\documentclass[preprint,12pt]{elsarticle}
\usepackage[utf8]{inputenc}
\usepackage{amsmath}
\usepackage{amsfonts}
\usepackage{amssymb}
\usepackage{graphicx}
\usepackage[running]{lineno}
\usepackage{setspace}
\usepackage{multirow}
\usepackage{booktabs}

\journal{Applied Mathematical Modelling}
\newtheorem{thm}{Theorem}

\newproof{pf}{\textbf{Proof}}

\begin{document}
\begin{frontmatter}
\title{Seasonality on the life cycle of Aedes aegypti mosquito and its effects on dengue outbreaks}

\author[label1]{Emilene  Pliego Pliego}
\ead{emilene.pliego@alumno.buap.mx}
\author[label1]{Jorge Vel\'azquez-Castro\corref{cor1}}
\ead{jorgevc@fcfm.buap.mx}
\author[label1]{Andr\'es Fraguela Collar}
\ead{fraguela@fcfm.buap.mx}
\cortext[cor1]{Corresponding author.}
\address[label1]{Facultad de Ciencias F\'isico-Matem\'aticas, Benem\'erita Universidad Aut\'onoma de Puebla, Avenida San Claudio y 18 Sur, Col. San Manuel, Puebla, M\'exico}

\begin{abstract}
Dengue is a vector-borne disease transmitted by the mosquito Aedes aegypti.
It has been observed that its incidence is strongly influenced by temperature and other abiotic factors like rainfall and humidity.

In this work we compare the effects of seasonality in temperature that affect the entomological parameters of the mosquito with precipitation that affects the hatchery capacity. We also analyze its joint action using a dynamical model for the life cycle of Aedes aegypti and historical weather data from 8 regions of Mexico.
We found that the joint action of different mechanisms can enhance the prevalence of the disease, but also inhibit it when they act in an asynchronous way.

We found that for the studied regions, the seasonality of the low temperature rather than mean temperature is the main driving force of Dengue outbreaks. We also analyzed the role of the diapause in these kinds of outbreaks.

The methodology developed here can be used to discover the underlying mechanism of Dengue outbreaks in different regions and thus help to apply targeted control measures. 
\end{abstract}

\begin{keyword}
Dengue, seasonality, Aedes, outbreak, diapause, model.
\end{keyword}

\end{frontmatter}

 \section{Introduction}
Dengue fever is the most important acute viral disease in humans transmitted by arthropods. The mosquito {\em Aedes aegypti} is the main transmitter of it. The etiologic agent is the ``Dengue virus'' (DENV) and has been pooled in four serotypes DENV-1, DENV-2, DENV-3, and DENV-4 all of them causing the disease \cite{Garcia}. The geographic distribution of mosquito Aedes is very wide. As a result of this, approximately 2.5 billion people are at risk of infection and 50 million infections occur annually \cite{OMS}.

Dengue disease is caused by the bite of an infected female {\em Aedes aegypti} mosquito \cite{Gordon}. Thus, it is thought that the main cause of seasonal dengue outbreaks is the increase of mosquito population during the outbreak period. In turn, the variability of the mosquito population can be caused by the alternation between favorable and unfavorable environmental conditions that affect the fitness of the mosquito population. These alternating conditions are produced by the seasonal weather patterns of a particular place.

There are many works, not only for Dengue but other mosquito borne diseases, that explain the seasonal patterns using entomological information. It has been found that small changes in the parameters can trigger changes in the dynamics of dengue disease \cite{Ferreira,rodriguesseasonality2014}. Also, Esteva and Yang introduced a model  that took into account the  temperature dependence on the entomological parameters to assess its effect on the incidence of dengue \cite{Esteva-Yang2015}.
Massad and Forattini calculated the expected increase in the density of female adult anopheline mosquitoes, assuming temperature-dependent functions on the rates of the mosquitoe's life cycles.
The main conclusion of this study, was that the increase in the density of adult females as a consequence of the temperature increase was related to an increase in malaria risk \cite{Massad}. In a similar way, there is some research that shows the sensitivity of mosquito Aedes aegypti to environmental factors like temperature and rainfall \cite{Dobson2006}
Furthermore, how temperature influences each stage of the mosquito's life cycle was studied by representing transitions and death rates as explicit functions of temperature \cite{Ferreira2}. 
In order to obtain the temperature dependence of the entomological parameters, laboratory experiments were performed and the results fitted to polynomials \cite{yangfollow2010, Yang2009b, Yang2009a}.

The temperature and its relation with the disease has been studied with probabilistic and dynamical models. On the other hand, other factors like precipitation and humidity have only been statistically analyzed. 
In addition to this, the period of suspended development with unfavorable conditions of a fraction of eggs, called diapause and its effect in Dengue outbreaks has just started to been considered \cite{adams}.

In this work we developed a dynamical model of the life cycle of the mosquito Aedes in order to study the differences that yearly weather cycles of precipitation, temperature and diapause generate in the seasonality of mosquito population. The entomological parameters in the model are temperature dependent and the hatchery capacity can vary with time according to the precipitation. We analyzed the predicted mosquito population for both seasonality in temperature and precipitation. We also studied their simultaneous action.
 Using weather and incidence data of 8 states of Mexico, we were able to obtain conclusions about the role of each mechanism in the generation of seasonality in dengue outbreaks for these regions.
Finally, we also studied the role of diapause in dengue outbreaks.

We were able to identify the relative importance of each mechanism in the studied regions. We also showed that even if the mosquito is dormant for a short period of time, the diapause mechanism can lead to its seasonal reemergence.

 \section{Formulation of the model}
 To assess weather seasonality as a dengue outbreak factor, we introduce a dynamical model of the {\em Aedes aegypti} mosquito life cycle . We consider  two main stages of the life cycle of Aedes, the mature mosquitoes (imago) and an aquatic stage (eggs).
We model just the population of the female mosquitoes $(m)$, as these are the only ones that can transmit Dengue, and $(p)$ represent the pre-adult mosquito population, specifically eggs.

\begin{figure}[h]\label{diagrama1}
\centering
\includegraphics[width= 150 pt]{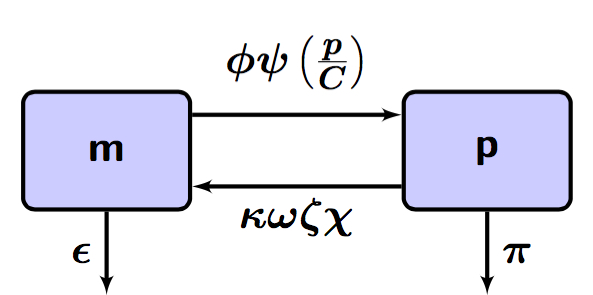}
\caption{Flowchart describing the life cycle of Aedes mosquitoes}
\end{figure}

	The oviposition rate is $\phi m \psi(p/C)$, where $\phi$ is the intrinsic oviposition rate at low densities, and $\psi$ is a continuous decreasing saturation function which satisfy $\psi([0,\ 1]) \subset [0,\ 1]$, $\psi(0)=1$ , $\psi(1)=0$ and $C$ represents the carrying capacity of the hatcheries.

	The female mosquito population $(m)$ increases accordingly with the per-capita development rate $\omega$, i.e. the rate at which a mosquito in its immature stage becomes adult, but it decreases accordingly with its per-capita mortality rate $\epsilon$. In addition to this, $\chi$ represents the fraction of larva that become pupa, $\zeta$ is the fraction of pupa that become adults, and $\kappa$ is the fraction of female newborn adult mosquitoes.
	The pre-adult mosquito population decreases according to the change of its own population into maturer stages and its death rate, $\omega$ and $\pi$, respectively. 
 The dynamic of both stages described above are modeled with the following differential equations

\begin{align*}\label{bloque2}
\frac{dm(t)}{dt}	&= \kappa\zeta\chi\omega p(t) - \epsilon m(t);\\
\frac{dp(t)}{dt}	&= \phi m(t)\psi\left(\frac{p(t)}{C}\right)- (\pi + \omega)p(t).& (I)
\end{align*}

\begin{table}[h]
	\centering\scalebox{.9}{
		\begin{tabular}{cl}
			\hline
			\textbf{Parameter}&	\textbf{Description}\\
			\hline
			$\phi$ & Oviposition rate.\\
			$\epsilon$ & Adult mosquito death rate.\\
			$\pi$  & Eggs death rate.\\
			$\omega$& Eggs maturation rate.\\
			$\kappa$ & Proportion of female mosquito in population.\\
			$\chi$ & Transition rate from larva to pupa.\\
			$\zeta$ & Transition rate from pupa to adult.\\
			\hline
 \end{tabular}}
 \caption{Parameters for the model of life cycle.}\label{tabla2}
\end{table}

\section{Analysis of the model}

For biological reasons, we need non negative solutions $m\geq 0$ and $0\leq p \leq C$, where $C$ is the carrying capacity of hatcheries. Thus, the region of biological interest where we study the solutions of the system $(I)$ is given by the set $\Omega = \{ (m,p)\in\mathbb{R}^{2}_{+}| \ m\geq 0 \ {\rm and} \ 0\leq p \leq C \}$.\\
We calculate the basic offspring number of mosquitoes, using the method \cite{Brauer} (see appendix A), that is:
\begin{equation}
R_{M}={\kappa\omega\zeta\chi\phi\over\epsilon(\pi+\omega)}
\end{equation}

This parameter will allow us to study the stability of the system, where $R_{M}$ is the average number of mosquitoes produced by a single mosquito during its lifespan. The biological meaning of this parameter is as follows: $\phi/\epsilon$ is the average number of eggs oviposited by one female mosquito, $1/(\pi+\omega)$ is the average time of survival of eggs, as such, then $(\kappa\omega\zeta\chi)(1/\pi+\omega)$ is the probability that an egg will survive to become a mature mosquito.  The product of the first and last quantities is equal to $R_{M}$ \cite{Esteva2006}. \\
We formulate a theorem that guarantees that the system $(I)$ is invariant on the $\Omega$ region.

\begin{thm}\label{teorema1}
	If the initial terms satisfy that $m(0)=m_{0}\geq0$, $p(0)=p_{0}\geq0$ and $m_{0}+p_{0}>0$, then the corresponding solution of system  $(I)$, $(m(t),p(t))$ satisfies that $m(t)>0$ and $p(t)>0$ for all $t>0$.\\
	Furthermore, for any initial condition $(m_{0}, p_{0})\in\Omega$, it holds that  $(m(t),p(t))\in\Omega$ for all $t>0$.
\end{thm}

\begin{pf}
To prove the first part of the theorem we multiply the first equation of $(I)$ by $p(t)$ and the second equation by $m(t)$ and add them,  we get

\begin{eqnarray*}
	p(t)\frac{dm(t)}{dt} + m(t)\frac{dp(t)}{dt}	&=& \kappa\omega\zeta\chi p^{2}(t) + \phi m^2(t)\psi\left(\frac{p(t)}{C}\right)- (\epsilon+ \pi + \omega)m(t)p(t).
\end{eqnarray*}

Therefore,
\begin{align*}
\frac{dm(t)p(t)}{dt} + (\epsilon+ \pi + \omega)m(t)p(t)	&= \kappa\omega\zeta\chi p^{2}(t) + \phi m^2(t)\psi\left(\frac{p(t)}{C}\right).
\end{align*}

Thus, due to  $m_{0}\geq 0$ and $p_{0}\geq 0$ which are unique because of the existence and uniqueness theorem, $\kappa\omega\zeta\chi p^{2}(t) + \phi m^2(t)\psi\left(\frac{p(t)}{C}\right)>0$. Since otherwise there would exist a
$t^{*}>0$ such that $p(t^{*})=m(t^{*})=0$ and $(m(t),p(t))$ would be the identically zero solution, then $p_{0}=m_{0}=0$ which contradicts the fact that $p_{0}+m_{0}>0$.

Then, we can conclude that

\begin{eqnarray*}
	\frac{dm(t)p(t)}{dt}	+ (\epsilon+ \pi + \omega)m(t)p(t)>0
\end{eqnarray*}so, $\frac{dm(t)p(t)}{m(t)p(t)}	>- (\epsilon+ \pi + \omega)dt$. That integrating  and using the initial conditions,  we get $m(t)p(t)>m_{0}p_{0}e^{-(\epsilon+\pi+\omega)t}\geq0$ for all $t>0$ , since $p_{0}\geq0 $ and $m_{0}\geq0$, thus $m(t)p(t)>0$ and due to $m(t)$ and $p(t)$ are continuous,
in this manner $m(t)>0$ and $p(t)>0$ whenever $t>0$.

For the second part of the proof we suppose that there is some $\tilde{t}$ such that $p(\tilde{t})=C$ and $\tilde{t_{1}}$ is the first value of  $t>0$ where $p(\tilde{t_{1}})=C$. Then for $t<\tilde{t_{1}}$ we get $p(t)<C$, so $p'(\tilde{t_{1}})> 0$. On the other hand, $p'(\tilde{t_{1}})=\phi m(\tilde{t_{1}})\psi(1)- (\pi + \omega)p(\tilde{t}_{1})=-(\pi + \omega)C<0$, that is a contradiction.
Then, we get that $p(t)<C$ for all $t>0$.
$\blacksquare$\\
\end{pf}

The system $(I)$ has two fixed points, the first one is the trivial fixed point $(0,0)$ and the second one is $(\tilde{m},\tilde{p})$ and satisfies the next equation.

\begin{align}\label{umbral2}
\psi\left(\frac{\tilde{p}}{C}\right) & =\frac{1}{R_{M}}, \\
\tilde{m} & =\frac{\kappa \omega \zeta\chi\tilde{p}}{\epsilon}\nonumber
\end{align}

\begin{thm}
 \begin{enumerate}
\item If $R_{M}>1$ the system $(I)$ admits two fixed points where:\\
	\begin{enumerate}	
		\item The trivial stationary solution $(0,0)$, is unstable and is a saddle point.
		\item The non trivial stationary solution $(\tilde{m},\tilde{p})$ is a locally asymptotically stable node.
	\end{enumerate}
\item If $0<R_{M}<1$ the system $(I)$ admits only $(0,0)$ as fixed point in $\Omega$ and is locally asymptotically stable node.
\end{enumerate}
\end{thm}

\begin{pf}
The stability properties of $(0,0)$ is given by the eigenvalues of the derivative of system $(I)$ evaluated in this point which is given by

\begin{equation*}
{\bf DF(0,0)}=\left(
\begin{array}{cc}
-\epsilon & \kappa\omega\zeta\chi\\
\\
\phi &  -(\pi + \omega)\\
\end{array}
\right).
\end{equation*}

Solving $Det({\bf \lambda\mathbb{I}- DF(0,0)})$ we get the following characteristic polynomial .

\begin{equation*}\label{caracteristica}
\lambda^{2}+ (\epsilon+\pi+\omega)\lambda + \epsilon(\pi+\omega)(1-R_{M})=0
\end{equation*}
whose roots are of the shape
\begin{equation}\label{root}
\lambda_{\pm}=\frac{1}{2}(\gamma\pm\sqrt{\xi})
\end{equation}
where $\gamma=-(\epsilon+\pi+\omega)$, $\xi=\gamma^{2}-4\chi$ y $\chi=\epsilon(\pi+\omega)(1- R_{M})$.\\

Note that $\gamma<0$ y $\xi=(\epsilon+\pi+\omega)^{2}- 4\epsilon(\pi+\omega)(1- R_{M})=\epsilon^{2}+ 2\epsilon(\pi+\omega)+(\pi+\omega)^{2} - 4\epsilon(\pi+\omega)+ 4\epsilon(\pi+\omega)R_{M} =((\pi+\omega)-\epsilon)^{2}+4\epsilon(\pi+\omega)R_{M}>0$.

\begin{itemize}
	\item If $R_{M}<1$, then $1-R_{M}>0$, so $\chi>0$ and $0<\xi$.
	Then we have two eigenvalues with the shape $\lambda_{-}<\lambda_{+}<0$, we can conclude that the stationary solution is a locally asymptotically stable node.
	\item If $R_{M}>1$, then  $1-R_{M}<0$, thus $\chi<0$ and $\xi>0$. Then $\lambda_{-}<0$ and $\lambda_{+}>0$, so the stationary solution is unstable and is called a saddle point.
\end{itemize}
	
In the same way the stability properties of  $(\tilde{m}, \tilde{p})$ is given by the eigenvalues of the derivative of system $(I)$ evaluated in this point which is given by

\begin{equation*}
{\bf DF(\tilde{m}, \tilde{p})}=\left(
\begin{array}{cc}
-\epsilon & \kappa\omega\zeta\chi\\
\\
\phi\psi(\frac{\tilde{p}}{C_{0}}) &  \frac{\phi}{C_{0}}\tilde{m}\psi^{'}(\frac{\tilde{p}}{C_{0}})-(\pi + \omega)\\
\end{array}
\right).
\end{equation*}

Solving $Det({\bf \lambda\mathbb{I}- DF(\tilde{m}, \tilde{p})})$ we get the following characteristic polynomial.
\begin{eqnarray*}\label{car2}
\lambda^{2}&+& \left(\epsilon+\pi+\omega -\frac{\phi}{C_{0}}\tilde{m}\psi^{'}\left(\frac{\tilde{p}}{C_{0}}\right)\right)\lambda - \frac{\epsilon\phi}{C_{0}}\tilde{m}\psi^{'}\left(\frac{\tilde{p}}{C_{0}}\right)\\ \nonumber
&+&  \epsilon(\pi+\omega)\left(1- R_{M}\psi\left(\frac{\tilde{p}}{C_{0}}\right)\right)=0
\end{eqnarray*}

Whose roots are of the shape \eqref{root} where
\begin{eqnarray*}
	\gamma&=&-(\epsilon+\pi+\omega-\frac{\phi}{C_{0}}\tilde{m}\psi^{'}\left(\frac{\tilde{p}}{C_{0}}\right))\\
	\chi&=&\epsilon(\pi+\omega)(1-R_{M}\psi\left(\frac{\tilde{p}}{C_{0}}\right))-\frac{\epsilon\phi}{C_{0}}\tilde{m}\psi^{'}\left(\frac{\tilde{p}}{C_{0}}\right)\\
	\xi&=&\gamma^{2}-4\chi
\end{eqnarray*}

As $\psi\left(\frac{\tilde{p}}{C_{0}}\right)=\frac{1}{R_{M}}$ then the above reduces to:
\begin{eqnarray*}
	\gamma&=&-(\epsilon+\pi+\omega-\frac{\phi}{C_{0}}\tilde{m}\psi^{'}\left(\frac{\tilde{p}}{C_{0}}\right))\\
	\chi&=&-\frac{\epsilon\phi}{C_{0}}\tilde{m}\psi^{'}\left(\frac{\tilde{p}}{C_{0}}\right)\\
	\xi&=&\gamma^{2}-4\chi
\end{eqnarray*}

Notice that $\gamma<0$ and  $\chi>0$ since a derivative of $\psi$ is a decreasing function.

\begin{eqnarray*}
\xi&=&\left[-\left(\epsilon+\pi+\omega-\frac{\phi}{C_{0}}\tilde{m}\psi^{'}\left(\frac{\tilde{p}}{C_{0}}\right)\right)\right]^{2}+4\frac{\epsilon\phi}{C_{0}}\tilde{m}\psi^{'}\left(\frac{\tilde{p}}{C_{0}}\right)\\
&=&\left[(\epsilon-(\pi+\omega))+ \frac{\phi}{C_{0}}\tilde{m}\psi^{'}\left(\frac{\tilde{p}}{C_{0}}\right)\right]^{2}+ 4\epsilon(\pi+\omega)
\end{eqnarray*}

Note that $\xi>0$ and  $\gamma<0$, so we get that $\lambda_{-}<\lambda_{+}<0$ and we can conclude that the stationary solution $(\tilde{m},\tilde{p})$ is a locally asymptotically stable node.$\blacksquare$\\ 
\end{pf}

\section{Is $R_{M}$ a good indicator of dengue risk?}

In order to perform numerical simulations, in what follows we will use
the explicit functional form $\psi(\frac{p}{C})=\left(1 - \frac{p}{C}\right)$. Thus the model becomes

\begin{align*}
\frac{dm(t)}{dt}	&= \kappa\omega\zeta\chi p(t) - \epsilon m(t);\\
\frac{dp(t)}{dt}	&= \phi m(t)\left(1-\frac{p(t)}{C}\right)- (\pi + \omega)p(t)& (II).
\end{align*}

All the conclusions for system $(I)$ are true for system $(II)$ as $(1-x)$ is a particular $\psi(x)$, but in this case we can obtain an explicit expression for the stationary points. In the case when $R_{M}<1$, the only stationary point with biological meaning is $(0,0)$ and it is an asymptotically stable node. This means that the adult and aquatic mosquito population will eventually die out. On the other hand, if $R_{M}>1$, there are two stationary points  $(0,0)$ and $(\tilde{m},\tilde{p})=\left(C\frac{\kappa\omega\zeta\chi}{\epsilon}\left(\frac{R_{M}-1}{R_{M}}\right), C\left(\frac{R_{M}-1}{R_{M}}\right)\right)$. The second one being the only asymptotically stable and the trivial one $(0,0)$ being unstable (saddle point). In this case, the mosquito population will eventually reach the nontrivial stationary point if either the mature or aquatic population is not initially zero.

The value of $R_{M}$ is a function of the entomological parameters  $\epsilon$, $\omega$, $\pi$, $\chi$, and $\zeta$. In turn , the entomological parameters are temperature dependent. Thus $R_{M}$ takes different values for different temperatures. In regions where the temperature is notably different from one season to another there is the possibility that $R_{M}$ is less than one for certain weeks during the year but greater than one for others. This is one of the main reasons $R_{M}$ does not give a good criterium to make a risk classification of a region.

One could be tempted to think that it is sufficient that $R_{M}<1$ at some point in the year in order for the mosquito die out. Thus, if the mosquito is not re-intoduced, this region would be a mosquito free region and thus dengue could not be spread. In this situation we have to take into account the time it takes the population of mosquitoes in air and aquatic phase to dies out. If this time is shorter than the period when $R_{M}<1$, then the mosquito dies out but this is not true if this time is larger than the period where $R_{M}<1$ is fulfilled.

In order to have an idea of the time of dying out and thus the necessary time that $R_{M}$ should be less than one to have a mosquito free region we perform numerical simulations.
We solve numerically the adimensional form of the system (II) by scaling with $C$. The initial conditions are between $0$ and $1$, with two sets of parameter values as reported in table (\ref{tabla}). The first set of values makes $R_{M}>1$ and the second one lead to $R_{M}<1$. In each simulation the time to reach a neighborhood to the stable stationary point was registered. 

\begin{table}[h]
\centering\scalebox{.9}{
\begin{tabular}{c|ccccccc}
\hline
\textbf{Parameter}	& $\phi$ & 	$\epsilon$ & $\pi$ & 	$\omega$ & $\kappa$ &  	$\chi$ & 	$\zeta$\\
\hline
\multirow{2}{1cm}{Values} & 10 & 0.05& 0.07& 0.05& 0.5 &  0.5 & 0.7 \\
& 10 & 0.05 & 0.07& 0.01& 0.5& 0.5& 0.7\\
\hline
\end{tabular}}
\caption{Values for each parameter.}
\label{tabla}
\end{table}

\begin{figure}
\centering
\includegraphics[width= 350 pt]{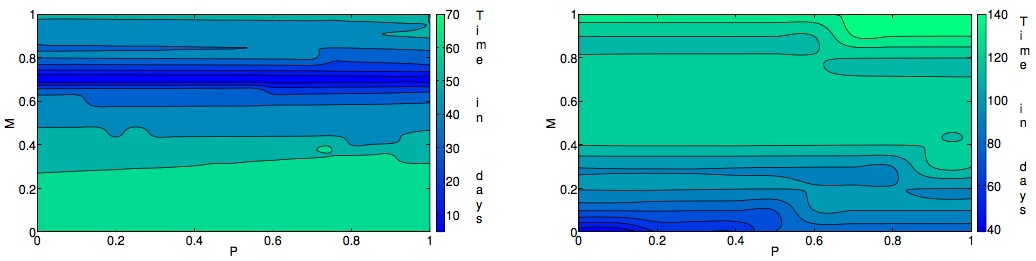}\label{figure4}
\caption{Time to reach the stable stationary point from different initial conditions. The neighborhood was taken with a size $0.001$.  {\bf Left:} $R_{M}>1$ 
{\bf Right:} $R_{M}<1$.}
\label{figuretime}
\end{figure}

We observe from Fig. \ref{figuretime} that a large set of initial conditions takes months to reach the stable stationary point. This time is of the order of a season, and thus for regions with marked weather seasonality we cannot expect that the system is near its equilibrium.

The most notorious consequence of this is that even if for a period of time $R_ {M} <1 $ and thus the vector population tend to extinction. It's extinction may take sufficient time so that the weather conditions may change before this happens and make $R_{M}>1$.  In this case the vector population will start to increase and if the cycle repeats, seasonality is observed.

This is a very important phenomena to take into account for risk evaluation, because it shows that even if the basic offspring number might be an indicator of the tendency of dengue risk transmission at a certain moment, it is not a good parameter to  build a static risk classification.

\section{Existence of periodic solutions}
If the system $(I)$ has periodic solutions, then seasonality in dengue outbreaks can be driven by this intrinsic periodicity of the system. In order to assess this possibility, periodical solutions are now discussed.

\begin{thm}\label{solperiodicas}
	For all $R_{M}>0$ the system $(I)$ does not have a periodic solution in the region of biological interest $\Omega$.
\end{thm}

\begin{pf}
Consider $F(m,p)=\kappa\omega\zeta\chi p-\epsilon m$ and $G(m,p)= \phi m\psi\left(\frac{p}{C}\right)-(\pi+\omega)p$ then
\begin{eqnarray*}
	\frac{\partial F}{\partial m}&=&-\epsilon\\
	\frac{\partial G}{\partial p}&=& \frac{\phi}{C} m\psi^{'}\left(\frac{p}{C}\right)-(\pi +\omega)
\end{eqnarray*}

On the other hand, we get $\frac{\partial F}{\partial p}=\kappa\omega\zeta\chi>0$ and $\frac{\partial G}{\partial m}=\phi \psi\left(\frac{p}{C}\right)>0$ which are positive. 
Since $\psi\left(\frac{p}{C}\right)$ is a decreasing function, then $\psi^{'}\left(\frac{p}{C_{0}}\right)<0$, therefore $\frac{\phi m}{C}\psi^{'}\left(\frac{p}{C_{0}}\right)\leq 0$ since $m\geq0$  and $-(\epsilon+\pi+\omega)<0$. Thus, we conclude that $\frac{\partial F}{\partial m}+\frac{\partial G}{\partial p}<0$ for all $(m,p)\in\Omega$. Therefore the system $(I)$ does not have a periodic solution in the region of biological interest  $\Omega$ due to Bendixon criteria (see \cite{Perko2001}). $\blacksquare$
\end{pf}

Hence the model predicts that seasonality in dengue outbreaks is not caused by intrinsic periodicity in the dynamics of the Ae. aegypti mosquito population.

\section{Identification of seasonality mechanism: Temperature vs Precipitation}

\subsection*{Variation in entomological parameters}

In order to analyze the effect of seasonality in temperature $T$, we now take into account the explicit dependence on $T$ of the entomological parameters according to \cite{yangfollow2010}.

\subsection{Variation in carrying capacity}

On the other hand, we propose that the carrying capacity of the hatcheries has a linear relation with the precipitation $P$ and is given by $C=C_{0} + b(\frac{P - \bar{P}}{\bar{P}})$ where $C_{0}$, $b<C_{0}$ are region specific constants and $\bar{P}$ is the average yearly precipitation per unit time. This relation models the deviation of the carrying capacity from its average value in a proportional way with the relative deviation of the precipitation from its yearly average.
Thus, the model predicts that in the rainy season the carrying capacity is greater than in the dry season.

\subsection{The seasonality}

Taking into account the variation of the parameters with temperature $T$ and precipitation $P$, the model is now a non autonomous system of equations,
\begin{align*}\label{modelo3}
\frac{dm(t)}{dt}	&= \kappa\omega(T)\zeta(T)\chi(T) p(t) - \epsilon(T) m(t);\\
\frac{dp(t)}{dt}	&= \phi(T) m(t)\psi\left(\frac{p(t)}{C(P)}\right)- (\pi(T) + \omega(T))p(t).& (III)
\end{align*}
where the temperature $T$ and the precipitation $P$ vary during the year.

We perform several numerical simulations of hypothetical situations to assess the effect of temperature and precipitation in the population of mosquitoes. We also analyze the joint action of seasonality in $T$ and $P$.

\begin{figure}[htbp]
\centering
\includegraphics[width= 350 pt]{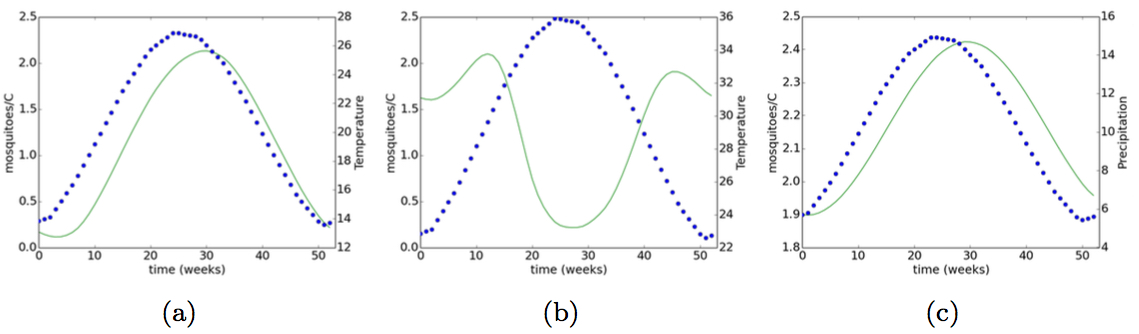}
	\caption{Seasonality of mosquito population generated by different mechanisms. solid line: mosquito population, dotted line: temperature in $(a)$ and $(b)$, precipitation in $(c)$. $(a)$ Temperature mechanism in temperate region ($T<30$ all year). $(b)$ Temperature mechanism in warm region ($T>30$ in hot season). $(c)$ Precipitation mechanism} \label{fig:figura4}
\end{figure}

In figure \ref{fig:figura4}.a we observe that in temperate regions, i.e. $T<30$ most of the time the increase in temperature lead to an increase in mosquito population but not always, as can be seen from the first 10 weeks of the simulated year. This is different from the effect of precipitation Fig. \ref{fig:figura4}.c which has the direct effect of always increasing the mosquito population if it rains more. The decreasing population even with increasing temperature at low temperatures is because during this period $R_{M}<1$. If this period just lasts a few weeks, the population may not become extinguished and at the moment when the temperature passes above a threshold value (around $15$ Celsius) where $R_{M}>1$ the mosquito population starts to grow again. In contrast to this mechanism in hot regions Fig. \ref{fig:figura4}.b, if temperature cross a high temperature threshold (around 29 Celsius), the fitness of the mosquito population starts to decrease gradually \cite{yangfollow2010}. This is the reason for the observed increment in population while the temperature is decreasing during week 29 to 45 in Fig. \ref{fig:figura4}.b. In general, yearly temperature variations lead to seasonality in the mosquito population but with different patterns whether it is a temperate or hot region. In turn each pattern is generated by a specific mechanism as explained above.

\begin{figure}[htbp!]
\centering
\includegraphics[width= 300 pt]{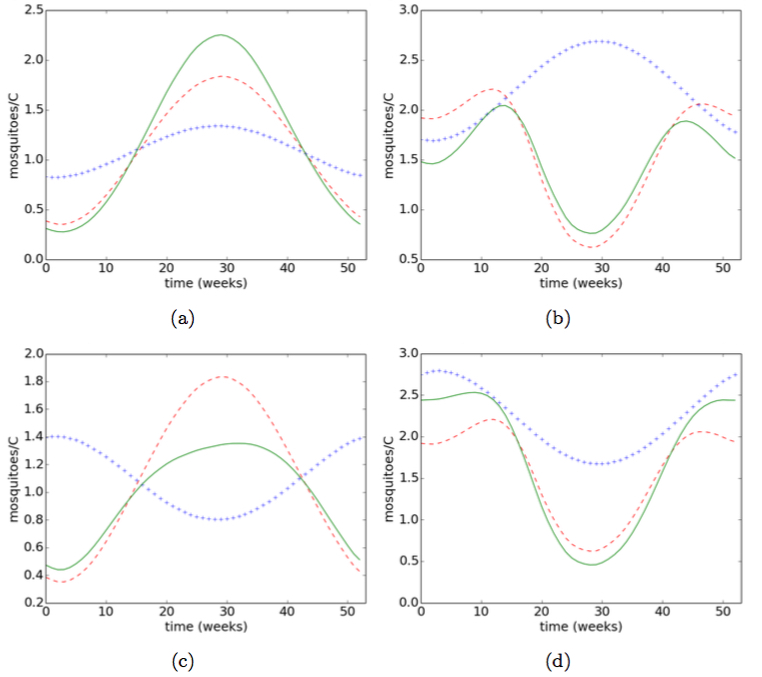}
\caption{Different seasonality patterns of mosquito population. Red dotted line: Temperature effect, blue dotted line: precipitation effect, continuous line: joint effect of temperature and precipitation. (a) synchronous hot and rainy seasons in temperate region, (b) synchronous hot and rainy seasons in warm region, (c) asynchronous hot and rainy seasons in temperate region, (d) asynchronous hot and rainy seasons in warm region.} \label{fig:figura5}
\end{figure}

Figure \ref{fig:figura5}.a shows that in a temperate region, the joint action of seasonality in precipitation and temperature enhance the seasonal mosquito outbreaks when acting in a synchronous way, i.e. when the rainy season coincides with the warm season. On the other hand, in a hot region the effect of the joint action is the other way around. This means that in hot regions if the warm season takes place at the same time of the rainy season, the mosquito outbreak is in a certain degree inhibited as can be seen in Fig \ref{fig:figura5}.b. This is because in hot regions precipitation and temperature have antagonistic effects on the mosquito population in contrast to temperate regions. As expected when the rainy and the warm season act asynchronously during the year, this effect is inverted as can be seen from Fig \ref{fig:figura5}.c and Fig \ref{fig:figura5}.d.

\begin{figure}
\centering
\includegraphics[width= 350 pt]{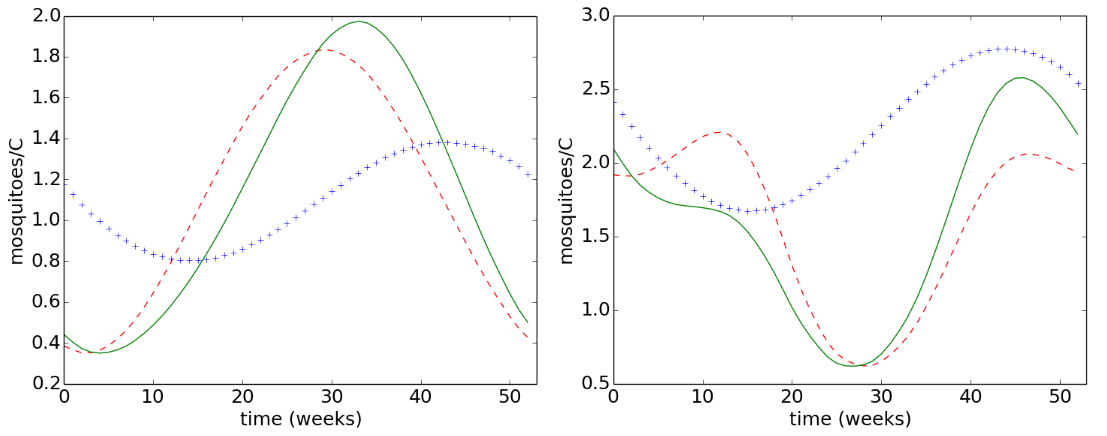}
\caption{Complex patterns in the dynamics of Ae. aegypti population. 
red dotted line: effect of temperature, blue dotted line: effect of precipitation, continuous line: joint effect of seasonality in temperature and precipitation. Left: temperate region, right: warm region.}
\label{figure4}
\end{figure}

If the rainy and warm season are not completely synchronous or asynchronous, then more complex patterns can emerge as shown in Fig. (\ref{figure4}).

\section{Diapause mechanism}

In certain places it has been observed that after a season without mosquitoes, they reappear in an outbreak.
A possible explanation for this is that in dry seasons, the hatcheries dry and the mature-mosquito population becomes extinct, these unfavorable conditions force some eggs into a dispause state where they can survive for long periods of time. In the next season, when there are more favorable conditions, these eggs continue to develop, allowing the mosquito population to grow again from apparent absence.
In order to verify and quantify the effect of diapause in the seasonal outbreaks we have to extend the model to include the population of eggs in diapause state. We will represent this population with $a$. There is a fraction of the eggs that belong to the pre-adult phase $p$ that enter in diapause stage if the hatchery gets dry. This happens with a rate that is proportional to the rate of change of the precipitation $\dot{P}$. The number of eggs that can enter in diapause state should also be proportional to the density of eggs in the hatcheries, i.e.  $\propto p/C$. In the same way, the number of eggs in dapause state that can return to a normal state are proportional to its density in the hatcheries $a/C$. Finally, there is a per-capita death rate of diapause eggs $\delta$. This model can be written with the following system of differential equations,

\begin{eqnarray*}\label{diapause_model}
 \frac{dm(t)}{dt}&=&\kappa\omega\zeta\chi p(t) - \epsilon m(t),\nonumber\\
 \frac{dp(t)}{dt}&=& \phi m(t)\psi\left(\frac{p(t)}{C}\right)- (\pi + \omega)p(t) + \alpha\frac{A}{C}\dot{P}, \hspace{.6cm} (IV) \\
  \frac{da(t)}{dt}&=&-\alpha\frac{A}{C}\dot{P} -\delta a(t),\nonumber\\
 A=p & if &  \dot{P}<0 \quad and\nonumber\\
 A=a & & \text{other way} \nonumber
 \end{eqnarray*}

\begin{figure}[htbp]
\centering
\includegraphics[width= 350 pt]{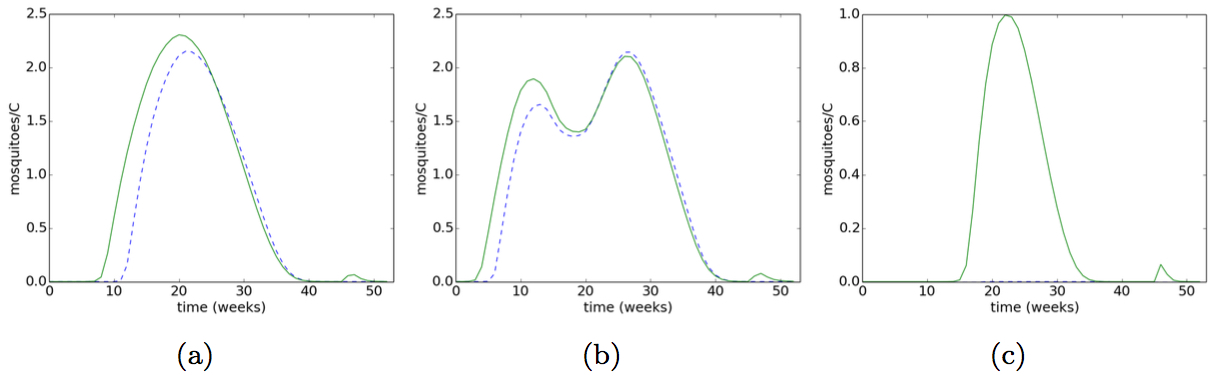}
\caption{Dynamics of mosquito population when diapause effect is present (solid line) and when it is not present (dashed line). $(a)$ temperate regions ($T<30$C all the seasons). $(b)$ Warm regions ($T>30$C in hot season). $(c)$ Cold season ($T<15$C) long enough to drive mature mosquitoes to extinction.} \label{fig:figura7}
\end{figure}

In Fig. \ref{fig:figura7} we see the differences in the mosquito population dynamics between a situation where diapause mechanism is present and a situation where it is absent. We were able to identify three possible situations, whether the region of interest is temperate, warm or cold. 

We observe from Fig. \ref{fig:figura7}.a that in temperate regions the diapause mechanism can enhance the mosquito outbreaks. The reason for this is because the eggs in diapuase state continue their development during the outbreak. In a warm region ($T>30$ in the hot season), there are two consecutive outbreaks as seen before. It is interesting to notice that diapause enhances just the first outbreak after the cold-dry season, see Fig. \ref{fig:figura7}.b. In the case when the cold season lasts long enough to drive mosquitoes to extinction there are two possibles outcomes. If there is no diapause phenomena, the mosquito goes extinct and as a consequence there is no seasonality the following years. On the other hand, if there is diapause mechanism, it allows the reappearance of the mosquitoes during the favorable seasons and thus seasonality appears \ref{fig:figura7}.c. The tiny outbreak shown around week 48 in all panels of Fig. \ref{fig:figura7} is another effect of the diapause phenomena.  The mechanism is as follows, the eggs in diapause stage continue its development when the rainy season starts. This allows new mature mosquitoes to develop, but at that time the temperature its still too cold for its effective survival. Thus, the mature mosquito population that comes from the eggs in diapause state die few days after birth.

\section{Study of 8 regions in Mexico}

We analyzed 8 regions with seasonal dengue outbreaks in Mexico in order to find the most important factors driving the dengue incidence seasonally.

We obtained dengue fever data  from January 2014 to December 2014 at state level from DGE-SINAVE \cite{dge}. The data includes all notified dengue cases. We focused our study in eight states of Mexico: Baja California Sur, Chiapas, Guerrero, Sonora, Colima, Oaxaca, Sinaloa, and Michoacan. We present the relation between incidence of dengue and mosquito population taking into account temperature and precipitation.

We suppose that the dengue transmission risk and thus dengue incidence have a linear relation with the vector population. With this assumption, a simple Pearson's correlation between forecasted mosquito population and dengue incidence allows us to test the main mechanism of the outbreaks.
We test the predictions with mean, low and high average weakly temperature in addition to precipitation.

\begin{table}[h]
\centering\scalebox{.75}{
\begin{tabular}{lcccccccc}
\hline
& \multicolumn{3}{c}{Temperature correlation} & & \multicolumn{3}{c}{Model correlation} \\ \cline{2-4} \cline{6-8}
Region & high & med & low & & high & med & low & time lag\\
B California Sur & 0.730422 & 0.764122 & 0.786509 & & 0.730854 & 0.595154 & 0.835484 & 4 \\
Chiapas & 0.522739 & 0.564111 & 0.681588 && 0.759529 & 0.562622 & 0.714105 & 5 \\
Guerrero & 0.576175 & 0.580331 & 0.579712 && 0.543532 & 0.481825 & 0.589716 & 5 \\
Sonora & 0.530424 & 0.495189 & 0.446617 && 0.533123 & 0.473598 & 0.841084 & 2 \\
Colima & 0.642097 & 0.676070 & 0.724862 && 0.610621 & 0.491214 & 0.689966 & 5 \\
Oaxaca & 0.619821 & 0.564582 & 0.606476 && 0.702489 & 0.563627 & 0.678785 & 4 \\
Sinaloa & 0.584191 & 0.507262 & 0.573519 && 0.640091 & 0.557325 & 0.646854 & 5
\end{tabular}}
\caption{Correlation between dengue incidence with temperature and between dengue incidence with the predicted mosquito population from the model.}
\label{tablas}
\end{table}

We found that the average weakly low temperature systematically has the higher correlation with dengue incidence (see Table \ref{tablas}), indicating that the low temperature rather than high or mean temperature is the one driving dengue outbreaks.
This is an important result because most of the studies so far use as the main data input the average temperature.

We also notice that the correlation with the incidence of the model is always higher than that with the temperature. This indicates that the model is actually giving more information than the bare temperature.

\begin{figure}[htbp]
\centering
\includegraphics[width= 350 pt]{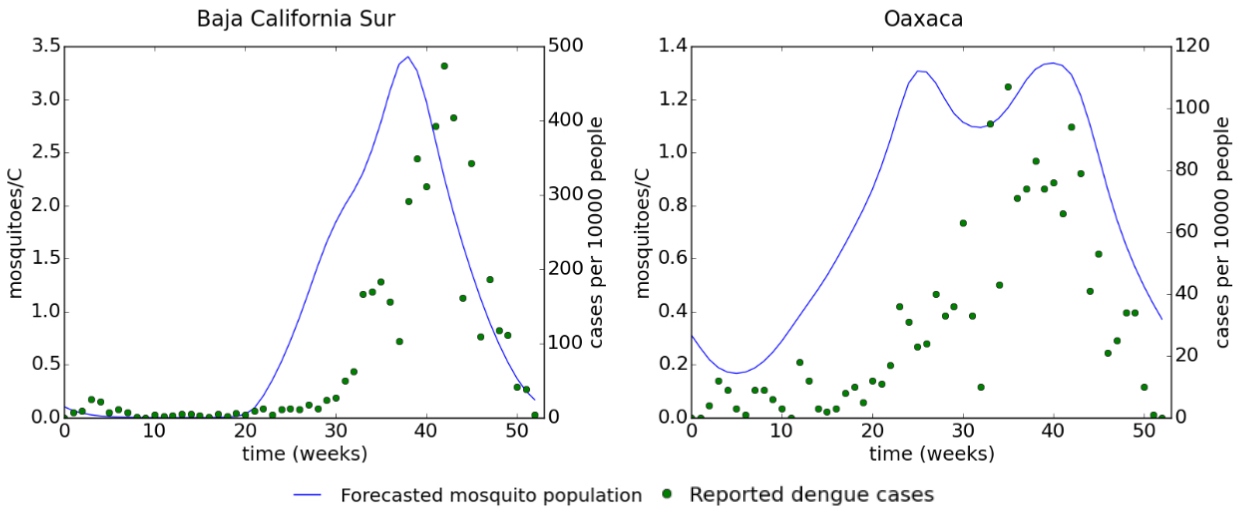}
\caption{Reported dengue cases in 2014 and forecasted mosquito population using the model.}\label{figurecalifoax}
\end{figure}

\begin{table}[tbh!]
\centering
\begin{tabular}{lcc|cc}
	& \multicolumn{2}{c}{Case a} &  \multicolumn{2}{c}{Case b} \\
		\toprule
	\textbf{Region}&	\textbf{Correlation}&	\textbf{Time lag}&	\textbf{Correlation}&	\textbf{Time lag}\\
	\midrule
	Baja California Sur & 0.944761 & 3& 0.839072 & 4\\
	Chiapas & 0.825987 & 3& 0.712882 & 5\\
	Guerrero & 0.607815 & 2& 0.599848 & 5\\
	Sonora & 0.673861 & 45 & 0.691131 & 13\\
	Colima & 0.750154 & 2 & 0.729818 & 3\\
	Oaxaca & 0.831021& 0 & 0.673992 & 4 \\
	Sinaloa & 0.755483 & 7& 0.647559 & 5\\
	Michoacan & 0.561575 & 39& 0.471490 & 41\\
	\bottomrule
\end{tabular}
\caption{ {{\bf Case a}: Correlation of forecasted mosquito population due to precipitation mechanism with incidence dengue data.{\bf Case b}: Correlation of forecasted mosquito population due to temperature mechanism with incidence dengue data. Time lag in weeks. }}\label{tablecorrelation1}
\end{table}

Comparing both columns of table (\ref{tablecorrelation1}) we can conclude that even if temperature alone is able to produce outbreak seasonality, precipitation plays a major role in the generation of seasonal outbreaks for most of the regions, i.e. there is a higher correlation of dengue incidence with mosquito populations increased by precipitation. On the other hand, in Sonora the main mechanism of seasonal outbreaks is the temperature effect. For the analyzed regions the rainy and warm seasons are synchronous. This is the reason why there is a high correlation with dengue incidence for both cases.

\begin{table}[tbh!]
\centering
\begin{tabular}{lcc|cc}
	& \multicolumn{2}{c}{Case a} &  \multicolumn{2}{c}{Case b} \\
		\toprule
	\textbf{Region}&	\textbf{Correlation}&	\textbf{Time lag}&	\textbf{Correlation}&	\textbf{Time lag}\\
	\midrule
		Baja California Sur & 0.920331 & 4& 0.914597 & 4\\
		Chiapas & 0.788383 & 4& 0.724022 & 7\\
		Guerrero & 0.633969 & 3& 0.630187& 7\\
		Sonora & 0.768424& 13 & 0.790213& 15\\
		Colima & 0.758208 & 3 & 0.734782 & 3\\
		Oaxaca & 0.759041 & 1 & 0.773997& 4 \\
		Sinaloa & 0.715874 & 7 & 0.699219 & 6 \\
		Michoacan & 0.471489 & 41& 0.774639 & 4\\
	\bottomrule
\end{tabular}
\caption{ {{\bf Case a}: Correlation of forecasted mosquito population due to precipitation and temperature mechanism with dengue incidences. {\bf Case b}: Correlation of forecasted mosquito population due to precipitation mechanism, temperature and diapause with dengue incidence. }}\label{tablecorrelation2}
\end{table}
	
It can be noticed from table (\ref{tablecorrelation2}) that in Guerrero and Sonora the correlation of the mosquito population with dengue incidence is higher when both mechanism, precipitation and temperature, are present. This is indicative that in Guerrero and Sonora the combined seasonality of precipitation and temperature is playing an important role in the observed pattern of dengue outbreaks.
\begin{figure}[htbp]
\centering
\includegraphics[width= 350 pt]{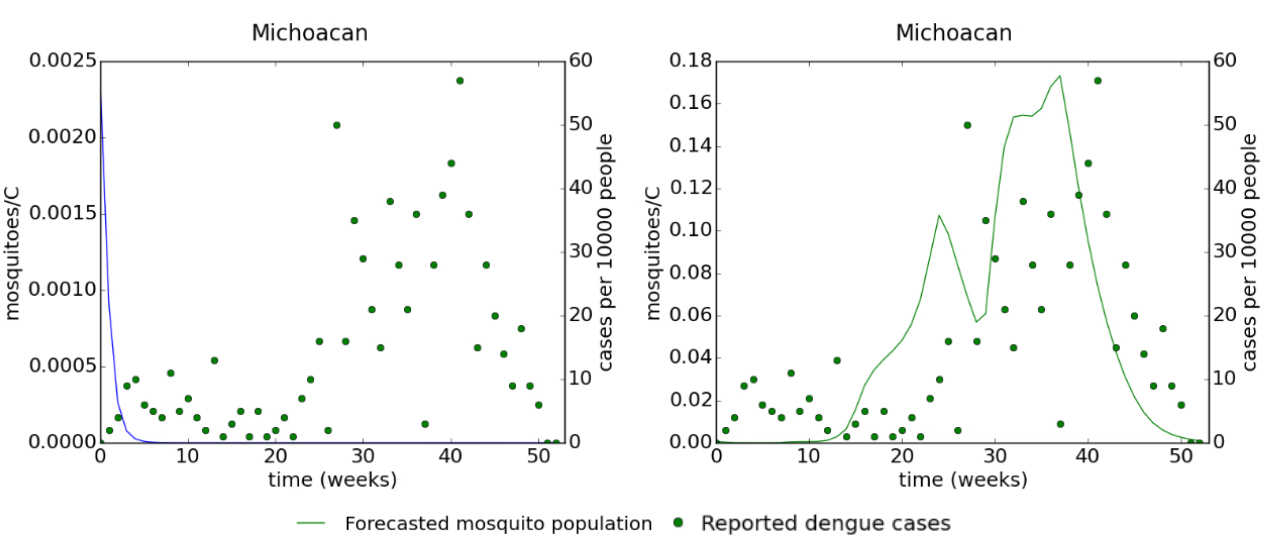}
\caption{Reported dengue cases and forecasted mosquito population for Michoacan state during 2014.
Left: Model prediction without considering diapause mechanism.
Right: Model prediction considering diapause mechanism.}\label{figuremichoacan}
\end{figure}

From table (\ref{tablecorrelation2}) we observe that the diapause mechanism is a very important factor in Michoacan. Without this mechanism there would be no dengue outbreaks as can be seen from Fig. (\ref{figuremichoacan}). This means that in this region the unfavorable conditions last long enough to drive the adult mosquito population to extinction but the eggs survive in a state of diapause. When the favorable weather conditions return, these eggs continue their development to adult mosquitoes and a dengue outbreak appears as a consequence of the mosquito population outbreak. Diapause is also enhancing the outbreaks in Sonora and Oaxaca.
	
 \section{Discussion}
 The seasonal dengue outbreaks can be attributed to the seasonality in mosquito population.
 This seasonality is mainly caused by the variation of the entomological parameters with temperature and the variation of carrying capacity of the hatcheries during the year. In certain circumstances the diapause mechanism is necessary to observe seasonal dengue outbreaks.
 There are seasons where the favorable temperature allows a faster development of the mosquito than in other seasons. The better fitness of the mosquito in some seasons is a consequence of a combination of faster reproduction rate and slower death rate between $25^{o}C - 30^{o}C$. 
 The differences in carrying capacity of the hatcheries between seasons is another factor that has a great influence on mosquitoes seasonality. This factor does not changes the basic reproductive number $R_{M}$ but it changes the maximum total population of the mosquitoes at a given moment.

 In some circumstances the time that the mosquito population reaches equilibrium is of the order of the length of a season. Thus we cannot evaluate $R_{M}$ in certain seasons and expect it to be a good indicator of a potential outbreak or not. This is so, because in a certain moment $R_{M}$ can be less than one and predict mosquito extinction, but if time to extinction is not reached in that season, the change of season with more favorable weather conditions will lead to a mosquito outbreak.

 This situation makes it necessary to follow the temperature dependence of the entomological parameters and the carrying capacity dependence in precipitation continuously.

For the studied region, the forecasted mosquito population using the model, follows the same seasonality as the dengue outbreaks. Thus we obtained high correlations between them. Furthermore, for the analyzed data the best correlation between mosquito population and dengue incidence is obtained when the entomological parameters are determined by the weekly average low temperature. This indicates that in regions where the temperature variation during the year is in the range of $5^{o}$C to $40^{o}$C, the cold weather is the limiting factor of the mosquito population, i.e. average low temperature rather than mean temperature is a better indicator of mosquito population and dengue transmission risk.

This important result should be taken into account in future studies because so far the average temperature is used rather than low temperature to parameterized predictive models. 
 
We have also shown that in some regions even if the unfavorable weather lasts long enough to drive the mosquito population to extinction, the diapause mechanism generates eggs reservoirs. This leads to the reappearance of mosquito outbreaks when favorable weather conditions reappear.

The studied weather factors, precipitation and temperature, lead to different patterns in the dynamics of the Ae. aegytpi and its joint action may result in more complex patterns. If the warm and rainy season coincide, the seasonal outbreaks are enhanced. The diapause mechanism also enhances the population growth when favorable conditions appear after a dry season.

We found that in Sonora the variation of the entomological parameters due to differences in season temperatures is the main mechanism causing seasonal dengue outbreaks. In the rest of studied regions, the more availability of hatcheries in the rainy season is the principal mechanism that leads to seasonal dengue outbreaks. This information helps to guide the control strategies, i.e. for the studied regions except for Sonora a suitable strategy would focus on the prevention of hatchery creation around 4 or 5 weeks before the dengue season. This study suggests that this same strategy would be less effective in Sonora. In Guerrero, Sonora and Colima the coincidence of the warm season with the rainy season is enhancing the outbreaks. While the diapause phenomena is allowing the seasonal outbreaks in Michoacan but it is also having an impact in Sonora and Oaxaca. 

\subsection{Conclusions}

We found that the joint action of seasonal precipitations and temperature can enhance an outbreak of Ae. aegypti population, but  also inhibit it when they act in an asynchronous way. Furthermore, an arbitrary time lag between rainy seasons and warm seasons can lead to complex patterns in the mosquito population and thus in dengue outbreaks.

We found that for the regions studied in Mexico, the seasonality of the low temperature rather than mean temperature is a more important factor driving Dengue outbreaks. The differences in average low temperature between seasons and the average precipitation of that season can be used to predict the dengue transmission risk via the Ae. aegipty population. 

On the other hand analysis of the patterns followed by the seasonal outbreaks can be used to determine the relative importance of the underlying mechanisms. This mechanism can be the increase of the basic reproduction number due to beneficial temperature, the growth of the hatchery capacity due to precipitation or the sudden development of eggs in diapause state. 

The methodology developed here can be used to discover the underling mechanism of dengue outbreaks in specific regions and thus help to apply control measures targeted to those places.

Regression Dengue risk models based in weather conditions need historical data to be calibrated. This work opens the possibility of making dengue risk assessment by forecasting regional weather without the need of a long time series of historical data.

\section{Acknowledgements} 
We are grateful for support from CONACyT and SEP-PRODEP Grant DSA/103.5/15/7449 (Mexican agencies)

 \section{References}
 \bibliographystyle{apalike}
 \bibliography{aedes-seasonality}

  \newpage
 \section*{Figure Captions}
 {\bf Figure 1:} Flowchart describing the life cycle of Aedes mosquitoes

 {\bf Figure 2:} Time to reach the stable stationary point from different initial conditions. The neighborhood was taken with a size $0.001$.  {\bf Left:} $R_{M}>1$ {\bf Right:} $R_{M}<1$.

{\bf Figure 3:} Seasonality of mosquito population generated by different mechanisms. solid line: mosquito population, dotted line: temperature in $(a)$ and $(b)$, precipitation in $(c)$. $(a)$ Temperature mechanism in temperate region ($T<30$ all year). $(b)$ Temperature mechanism in warm region ($T>30$ in hot season). $(c)$ Precipitation mechanism

{\bf Figure 4:} Different seasonality patterns of mosquito population. Red dotted line: Temperature effect, blue dotted line: precipitation effect, continuous line: joint effect of temperature and precipitation. (a) synchronous hot and rainy seasons in temperate region, (b) synchronous hot and rainy seasons in warm region, (c) asynchronous hot and rainy seasons in temperate region, (d) asynchronous hot and rainy seasons in warm region.

{\bf Figure 5:} Complex patterns in the dynamics of Ae. aegypti population. 
red dotted line: effect of temperature, blue dotted line: effect of precipitation, continuous line: joint effect of seasonality in temperature and precipitation. Left: temperate region, right: warm region.

{\bf Figure 6:} Dynamics of mosquito population when diapause effect is present (solid line) and when it is not present (dashed line). $(a)$ temperate regions ($T<30$C all the seasons). $(b)$ Warm regions ($T>30$C in hot season). $(c)$ Cold season ($T<15$C) long enough to drive mature mosquitoes to extinction.

{\bf Figure 7:} Reported dengue cases in 2014 and forecasted mosquito population using the model.

{\bf Figure 8:} Reported dengue cases and forecasted mosquito population for Michoacan state during 2014. Left: Model prediction without considering diapause mechanism.
Right: Model prediction considering diapause mechanism.

 \newpage
 \appendix
 \section{Basic offspring number}
Using the method \cite{Brauer}, we calculate the basic offspring number of mosquitoes, we write the system $(I)$ as $\mathfrak{\dot{x}}=\mathfrak{F}-\mathfrak{V}$.

 \begin{equation*}
 {\mathfrak{\dot{x}}}=\left(
 \begin{array}{cc}
 \dot{m}\\
 \\
 \dot{ p}\\
 \end{array}
 \right),
 \ \
 {\mathfrak{F}}=\left(
 \begin{array}{cc}
 \kappa\omega\zeta\chi p\\
 \\
 0\\
 \end{array}
 \right),
 \ \
 {\mathfrak{V}}=\left(
 \begin{array}{cc}
 \epsilon m\\
 \\
 (\pi + \omega)p-\phi m\psi\left(\frac{p}{C}\right)\\
 \end{array}
 \right).
 \end{equation*}
 The Jacobian matrices $F$ and $V$, associated with $\mathfrak{F}$ and $\mathfrak{V}$ respectively,
 at the mosquitoes free equilibrium point $m=0$, $p=0$ are.

 \begin{equation*}
 {F}=\left(
 \begin{array}{cc}
 0&\kappa\omega\zeta\chi \\
 \\
 0&0\\
 \end{array}
 \right),
 \ \
 {V}=\left(
 \begin{array}{cc}
 \epsilon & 0\\
 \\
 -\phi & (\pi + \omega)\\
 \end{array}
 \right),
 \ \
 {V^{-1}}=\left(
 \begin{array}{cc}
 \frac{1}{\epsilon} & 0\\
 \\
 \frac{\phi}{\epsilon(\pi + \omega)} & \frac{1}{\pi + \omega}\\
 \end{array}
 \right),
 \end{equation*}

 \begin{equation*}
 {K}=FV^{-1}=\left(
 \begin{array}{cc}
 \frac{\kappa\omega\zeta\chi\phi}{\epsilon(\pi + \omega)} & \frac{\kappa\omega\zeta\chi}{\pi + \omega}\\
 \\
 0 & 0\\
 \end{array}
 \right).
 \end{equation*}

 The eigenvalues of $K$ are $0$ and $\frac{\kappa\omega\zeta\chi\phi}{\epsilon(\pi + \omega)}$, so the basic offspring number is given by:
 \begin{equation}
 R_{M}={\kappa\omega\zeta\chi\phi\over\epsilon(\pi+\omega)}
 \end{equation}

 \end{document}